# ON THE BIT-COMPLEXITY OF
# SPARSE POLYNOMIAL MULTIPLICATION

*Joris van der Hoeven*

Laboratoire de Mathématiques
UMR 8628 CNRS
Université Paris-Sud
91405 Orsay Cedex
France

*Email:* `joris@texmacs.org`
*Web:* `http://www.math.u-psud.fr/~vdhoeven`

*Grégoire Lecerf*

Laboratoire de Mathématiques
UMR 8100 CNRS
Université de Versailles
45, avenue des États-Unis
78035 Versailles Cedex
France

*Email:* `Gregoire.Lecerf@math.uvsq.fr`
*Web:* `http://www.math.uvsq.fr/~lecerf`

*January 26, 2009*

In this paper, we present fast algorithms for the product of two multivariate polynomials in sparse representation. The bit complexity of our algorithms are studied in detail for various types of coefficients, and we derive new complexity results for the power series multiplication in many variables. Our algorithms are implemented and freely available within the MATHEMAGIX software. We show that their theoretical costs are well-reflected in practice.

A.M.S. SUBJECT CLASSIFICATION: 68W30, 12-04, 30B10, 42-04, 11Y05

## 1. INTRODUCTION

It is classical [SS71, Sch77, CK91, Für07] that the product of two integers, or two univariate polynomials over any ring, can be performed in *softly linear time* for the usual dense representations. More precisely, two integers of bit-size at most $n$ can be multiplied in time $\tilde{O}(n) = n (\log n)^{O(1)}$ on a Turing machine, and the product of two univariate polynomials can be done with $\tilde{O}(n)$ arithmetic operations in the coefficient ring.

Multivariate polynomials often admit many zero coefficients, so a *sparse representation* is usually preferred over a dense one: each monomial is a pair made of a coefficient and an exponent written in the dense binary representation. A natural problem is whether the product $R = PQ$ of two sparse multivariate polynomials in $z_1, ..., z_n$ can be computed in softly linear time. We will assume to be given a subset $X \subseteq \mathbb{N}^n$ of size $s$ that contains the support of $R$. We also let $d_1, ..., d_n \in \mathbb{N}$ be the minimal numbers with $X \subseteq \prod_{j=1}^n \{0, ..., d_j - 1\}$.

For coefficient fields of characteristic zero, it is proved in [CKL89] that $R$ can be computed using $\tilde{O}(s)$ operations over the ground field. This algorithm uses fast evaluation and interpolation at suitable points built from prime numbers. Unfortunately, the method hides an expensive growth of intermediate integers involved in the linear transformations, which prevents the algorithm from being softly linear in terms of bit-complexity.

In this paper, we turn our attention to various concrete coefficient rings for which it makes sense to study the problem in terms of bit-complexity. For these rings, we will present multiplication algorithms which admit softly linear bit-complexities, under the mild assumption that $\log d = (\log s)^{O(1)}$. Our approach is similar to [CKL89], but relies on a different kind of evaluation points. Furthermore, finite fields form an important special case for our method.

Let us briefly outline the structure of this paper. In section 2, we start with a presentation of the main algorithm over an arbitrary effective algebra $\mathbb{A}$ with elements of sufficiently high order. In section 3, we treat various specific coefficient rings. In section 4 we give an application to multivariate power series multiplication. In the last section 5, we report on timings with our MATHEMAGIX implementation [vdH+02b].

## 2. SPARSE POLYNOMIAL MULTIPLICATION

### 2.1. General setting

Let $\mathbb{A}$ be an effective algebra over an effective field $\mathbb{K}$, i.e. all algebra and field operations can be performed by algorithm.

We will denote by $\mathsf{M}(n) = \mathsf{M}_{\mathbb{K}}(n)$ the cost for multiplying two univariate polynomials of degree $n$ over $\mathbb{K}$ in terms of the number of arithmetic operations in $\mathbb{K}$. Similarly, we denote by $\mathsf{I}(n)$ the time needed to multiply two integers of bit-size at most $n$. One can take $\mathsf{M}(n) = O(n \log n \log \log n)$ [CK91] and $\mathsf{I}(n) = O(n \log n \, 2^{\log^* n})$ [Für07], where $\log^*$ represents the iterated logarithm of $n$. Throughout the paper, we will assume that $\mathsf{M}(n)/n$ and $\mathsf{I}(n)/n$ are increasing. We also assume that $\mathsf{M}(O(n)) = O(\mathsf{M}(n))$ and $\mathsf{I}(O(n)) = O(\mathsf{I}(n))$.

Given a multivariate polynomial $P \in \mathbb{A}[z] = \mathbb{A}[z_1, ..., z_n]$ and an index $i \in \mathbb{N}^n$, we denote $z^i = z_1^{i_1} \cdots z_n^{i_n}$ and let $P_i$ be the coefficient of $z^i$ in $P$. The *support* of $P$ is defined by $\operatorname{supp} P = \{i \in \mathbb{N}^n : P_i \neq 0\}$ and we denote by $s_P = |\operatorname{supp} P|$ its cardinality.

In the sparse representation, the polynomial $P$ is stored as a sequence of exponent-coefficient pairs $(i, P_i) \in \mathbb{N}^n \times \mathbb{A}$. Natural numbers are represented by their sequences of binary digits. The *bit-size* of an exponent $i \in \mathbb{N}^n$ is $l_i = \sum_{j=1}^n l_{i_j}$, where $l_{i_j} = \lceil \log_2 (i_j + 1) \rceil$. We let $e_P = \sum_{i \in \operatorname{supp} P} l_i$ be the bit-size of $\operatorname{supp} P$.

In this and the next section, we are interested in the multiplication $R = P Q$ of two sparse polynomials $P, Q \in \mathbb{A}[z]$. We assume given a finite set $X \subseteq \mathbb{N}^n$, such that $X \supseteq \operatorname{supp} R$. We will write $s = |X|$ for its size and $e = \sum_{i \in X} l_i$ for its bit-size. We also denote $\sigma = s_P + s_Q + s$ and $\epsilon = e_P + e_Q + e$. For each $j$, we introduce $d_j = \min \{k \in \mathbb{N} : \exists i \in \mathbb{N}^n, i_j = k, z^i \in X\} + 1$, which sharply bounds the partial degree in $z_j$ for each monomial in $X$. We denote $d = d_1 \cdots d_n > 0$.

## 2.2. Evaluation, interpolation and transposition

Given $t$ pairwise distinct points $\omega_0, ..., \omega_{t-1} \in \mathbb{K}^*$ and $s \in \mathbb{N}$, let $E: \mathbb{A}^s \to \mathbb{A}^t$ be the linear map which sends $(a_0, ..., a_{s-1})$ to $(A(\omega_0), ..., A(\omega_{t-1}))$, with $A = a_{s-1} u^{s-1} + \cdots + a_0$. In the canonical basis, this map corresponds to left multiplication by the generalized Vandermonde matrix

$$V_{s,\omega_0,...,\omega_{t-1}} = \begin{pmatrix} 1 & \omega_0 & \cdots & \omega_0^{s-1} \\ 1 & \omega_1 & \cdots & \omega_1^{s-1} \\ \vdots & \vdots & & \vdots \\ 1 & \omega_{t-1} & \cdots & \omega_{t-1}^{s-1} \end{pmatrix}.$$

The computation of $E$ and its inverse $E^{-1}$ (if $t = s$) correspond to the problems of multi-point evaluation and interpolation of a polynomial. Using binary splitting, it is classical [MB72, Str73, BM74] that both problems can be solved in time $O(\lceil t/s \rceil \mathsf{M}(s) \log s)$. Notice that the algorithms only require vectorial operations in $\mathbb{A}$ (additions, subtractions and multiplications with elements in $\mathbb{K}$).

Our main algorithm relies on the efficient computations of the transpositions $E^\top$, $(E^{-1})^\top: (\mathbb{A}^t)^* \to (\mathbb{A}^s)^*$ of $E$ and $E^{-1}$. The map $E^\top$ corresponds to left multiplication by

$$V_{s,\omega_0,...,\omega_{t-1}}^\top = \begin{pmatrix} 1 & 1 & \cdots & 1 \\ \omega_0 & \omega_1 & \cdots & \omega_{t-1} \\ \vdots & \vdots & & \vdots \\ \omega_0^{s-1} & \omega_1^{s-1} & \cdots & \omega_{t-1}^{s-1} \end{pmatrix}.$$

By the transposition principle [Bor56, Ber], the operations $E^\top$ and $(E^{-1})^\top$ can again be computed in time $O(\lceil t/s \rceil \mathsf{M}(s) \log s)$.

There is an efficient direct approach for the computation of $E^\top$ [BLS03]. Given a vector $a \in (\mathbb{A}^t)^*$ with entries $a_0, ..., a_{t-1}$, the entries $b_0, ..., b_{s-1}$ of $E^\top(a)$ are identical to the first $s$ coefficients of the power series

$$\sum_{i<t} \frac{a_i}{1 - \omega_i u}.$$

The numerator and denominator of this rational function can be computed using binary splitting. If $t \leqslant s$, then this requires $O(\mathsf{M}(s) \log s)$ vectorial operations in $\mathbb{A}$ [GG02, Theorem 10.10]. The truncated division of the numerator and denominator at order $s$ requires $O(\mathsf{M}(s))$ vectorial operations in $\mathbb{A}$. If $t > s$, then we cut the sum into $\lceil t/s \rceil$ parts of size $\leqslant s$, and obtain the complexity bound $O(\lceil t/s \rceil \mathsf{M}(s) \log s)$.

Inversely, assume that we wish to recover $a_0, ..., a_{s-1}$ from $b_0, ..., b_{s-1}$, when $t = s$. For simplicity, we assume that the $\omega_i$ are non-zero (this will be the case in the sequel). Setting $B(u) = b_{s-1} u^{s-1} + \cdots + b_0$, $D(u) = (1 - \omega_0 u) \cdots (1 - \omega_{s-1} u)$ and $S = B D$, we notice that $S(\omega_i^{-1}) = -a_i (u D')(\omega_i^{-1})$ for all $i$. Hence, the computation of the $a_i$ reduces to two multi-point evaluations of $S$ and $-u D'$ at $\omega_0^{-1}, ..., \omega_{s-1}^{-1}$ and $s$ divisions. This amounts to a total of $O(\mathsf{M}(s) \log s)$ vectorial operations in $\mathbb{A}$ and $O(s)$ divisions in $\mathbb{K}$.

## 2.3. General multiplication algorithm

Let $u$ be a new variable. We introduce the vector spaces

$$\begin{aligned} \mathcal{A} &= \{A \in \mathbb{A}[z] : \forall j, \deg_{z_j} A < d_j\} \\ \mathcal{B} &= \{B \in \mathbb{A}[u] : \deg B < d\} \end{aligned}$$

Given $i \in \mathbb{N}^n$, let $\kappa(i) = i_1 + i_2 d_1 + \cdots + i_n d_1 \cdots d_{n-1}$. The *Kronecker isomorphism* $\mathrm{K}: \mathcal{A} \to \mathcal{B}$, is the unique $\mathbb{K}$-linear map with

$$\mathrm{K}(z_1^{i_1} \cdots z_n^{i_n}) = u^{\kappa(i)}, \quad \text{for all } i_1 < d_1, ..., i_n < d_n.$$

It corresponds to the evaluation at $z = (u, u^{d_1}, ..., u^{d_1 \cdots d_{n-1}})$, so that $\mathrm{K}(R) = \mathrm{K}(P) \mathrm{K}(Q)$.

Assume now that we are given an element $\omega \in \mathbb{K}$ of multiplicative order at least $d$ and consider the following evaluation map

$$\begin{aligned} \mathrm{E}: \mathbb{A}[z] &\longrightarrow \mathbb{A}^s \\ A &\longmapsto (\mathrm{K}(A)(1), \mathrm{K}(A)(\omega), ..., \mathrm{K}(A)(\omega^{s-1})). \end{aligned}$$

We propose to compute $R$ though the equality $\mathrm{E}(R) = \mathrm{E}(P) \mathrm{E}(Q)$.

Given $Y = \{i_1, ..., i_t\} \subseteq \prod_{j=1}^n \{0, ..., d_j - 1\}$, let $V_{s,Y,\omega}$ be the matrix of $\mathrm{E}$ restricted to the space of polynomials with support included in $Y$. Setting $k_j = \kappa(i_j)$, we have

$$V_{s,Y,\omega} := V_{s,\omega^{k_1},...,\omega^{k_t}}^\top = \begin{pmatrix} 1 & 1 & \cdots & 1 \\ \omega^{k_1} & \omega^{k_2} & \cdots & \omega^{k_t} \\ \vdots & \vdots & & \vdots \\ \omega^{(s-1)k_1} & \omega^{(s-1)k_2} & \cdots & \omega^{(s-1)k_t} \end{pmatrix}.$$

Taking $Y = \operatorname{supp} P$ resp. $Y = \operatorname{supp} Q$, this allows us to compute $\mathrm{E}(P)$ and $\mathrm{E}(Q)$ using our algorithm for transposed multi-point evaluation from section 2.2. We obtain $\mathrm{E}(R)$ using one Hadamard product $\mathrm{E}(P) \mathrm{E}(Q)$. Taking $Y = X$, the points $\omega^{k_1}, ..., \omega^{k_t}$ are pairwise distinct, since the $k_j$ are smaller than the order of $\omega$. Hence $V_{s,X,\omega}$ is invertible and we recover $R$ from $\mathrm{E}(R)$ using transposed multi-point interpolation.

THEOREM 1. *Given two polynomials $P$ and $Q$ in $\mathbb{A}[z_1, ..., z_n]$ and an element $\omega \in \mathbb{K}$ of order at least $d$, then the product $PQ$ can be computed using $O(\epsilon)$ products in $\mathbb{K}$, $O(s)$ inversions in $\mathbb{K}$, $O(s)$ products in $\mathbb{A}$, and $O\bigl(\frac{\sigma}{s} \mathsf{M}(s) \log s\bigr)$ vectorial operations in $\mathbb{A}$.*

PROOF. By classical binary powering, the computation of the sequence $\omega^{d_1}, ..., \omega^{d_1 \cdots d_{n-1}}$ takes $O(e)$ operations in $\mathbb{K}$ because each $d_j - 1$ does appear in the entries of $X$. Then the computation of all the $\omega^{\kappa(i)}$ for $i \in \operatorname{supp} P$ (resp. $\operatorname{supp} Q$ and $X$) requires $O(e_P)$ (resp. $O(e_Q)$ and $O(e)$) products in $\mathbb{K}$. Using the complexity results from section 2.2, we may compute $\mathrm{E}(P)$ and $\mathrm{E}(Q)$ using $O((\lceil s_P/s \rceil + \lceil s_Q/s \rceil) \mathsf{M}(s) \log s)$ vectorial operations in $\mathbb{A}$. We deduce $\mathrm{E}(R)$ using $O(s)$ more multiplications in $\mathbb{A}$. Again using the results from section 2.2, we retrieve the coefficients $R_i$ after $O(\mathsf{M}(s) \log s)$ further vectorial operations in $\mathbb{A}$ and $O(s)$ divisions in $\mathbb{K}$. Adding up the various contributions, we obtain the theorem. □

REMARK 2. Similar to FFT multiplication, our algorithm falls into the general category of multiplication algorithms by evaluation and interpolation. This makes it possible to work in the so-called "transformed model" for several other operations besides multiplication.

## 3. VARIOUS COEFFICIENT RINGS

### 3.1. Finite fields

If $\mathbb{K}$ is the finite field $\mathbb{F}_{p^k}$ with $p^k$ elements, then its multiplicative group is cyclic of order $p^k - 1$. Whenever $p^k - 1 \geqslant d$, it follows that the main theorem 1 applies for any primitive element $\omega$ of this group.

Usually, $\mathbb{F}_{p^k}$ is given as the quotient $\mathbb{F}_p[u]/G(u)$ for some monic and irreducible polynomial $G$ of degree $k$. In that case, a multiplication in $\mathbb{K}$ amounts to $O(\mathsf{M}_{\mathbb{F}_p}(k))$ ring operations in $\mathbb{F}_p$. An inversion in $\mathbb{K}$ requires an extended gcd computation in $\mathbb{F}_p[u]$ and gives rise to $O(\mathsf{M}_{\mathbb{F}_p}(k)\log k)$ operations in $\mathbb{F}_p$. Using Kronecker multiplication, we can also take $\mathsf{M}_{\mathbb{K}}(n) = O(\mathsf{M}_{\mathbb{F}_p}(n\,k))$. Using these estimates, Theorem 1 implies:

COROLLARY 3. *Assume $p^k - 1 \geqslant d$. Given two polynomials $P$ and $Q$ in $\mathbb{F}_{p^k}[z_1, ..., z_n]$, the product $PQ$ can be computed using*

$$O\Big(\frac{\sigma}{s}\,\mathsf{M}_{\mathbb{F}_p}(s\,k)\log s + (s\log k + \epsilon)\,\mathsf{M}_{\mathbb{F}_p}(k)\Big)$$

*ring operations in $\mathbb{F}_p$ and $O(s)$ inversions in $\mathbb{F}_p$.*

Applying the general purpose algorithm from [CK91], two polynomials of degree $n$ over $\mathbb{F}_p$ can be multiplied in time $O(\mathsf{I}(\log p)\,n\log n\log\log n)$. Alternatively, we may lift the multiplicants to polynomials in $\mathbb{Z}[u]$, use Kronecker multiplication and reduce modulo $p$. As long as $\log n \in O(\log p)$, this yields the better complexity bound $O(\mathsf{I}(n\log p))$. Theorem 1 therefore implies:

COROLLARY 4. *Assume $p^k - 1 \geqslant d$ and $\log(s\,k) \in O(\log p)$. Given two polynomials $P$ and $Q$ in $\mathbb{F}_{p^k}[z_1, ..., z_n]$, the product $PQ$ can be computed in time*

$$O\Big(\frac{\sigma}{s}\,\mathsf{I}(s\,k\log p)\log s + (s\log k + \epsilon)\,\mathsf{I}(k\log p) + s\,\mathsf{I}(\log p)\log\log p\Big).$$

REMARK 5. *If $p^k - 1 < d$ then it is always possible to build an algebraic extension of suitable degree $l$ in order to apply the corollary. Such constructions are classical, see for instance [GG02, Chapter 14]. We need to have $p^{kl} - 1 \geqslant d$, so $l$ should be taken of the order $\log_{p^k} d$, which also corresponds to the additional overhead induced by this method.*

REMARK 6. *Under the generalized Riemann hypothesis, a primitive element in $\mathbb{F}_{p^k}$ can be constructed in polynomial time [BS91]. If $p$ is odd, then $\omega$ is a primive element if and only if $\omega^{(p^k-1)/2} = -1$. In $\mathbb{F}_p$, the smallest $\omega_{\mathrm{sm}} \in \mathbb{N}$ such that $\omega \bmod p$ is a primitive element satisfies $\omega_{\mathrm{sm}} = O((\log p)^6)$.*

## 3.2. Integer coefficients

### 3.2.1. Big prime algorithm

One approach for the multiplication $R = PQ$ of polynomials with integer coefficients is to reduce the problem modulo a suitable prime number $p$. This prime number should be sufficiently large such that $R$ can be read off from $R \bmod p$ and such that $\mathbb{F}_p$ admits elements of order $\geqslant d$.

Let $l_P = \max_i l_{|P_i|}$ denote the maximal bit-length of the coefficients of $P$ and similarly for $Q$ and $R$. Since

$$\max_i |R_i| \leqslant \min(s_P, s_Q) \max_i |P_i| \max_i |Q_i|,$$

we have

$$l_R \leqslant l := l_P + l_Q + \log_2 \min(s_P, s_Q).$$

It therefore suffices to take $p > \max(2^{l+1}, d)$. Corollary 4 now implies:

COROLLARY 7. *Given $P, Q \in \mathbb{Z}[z_1, ..., z_n]$ and a prime number $p > \max{(2^{l+1}, d)}$, we can compute $PQ$ in time*

$$O\Big(\frac{\sigma}{s} \,\mathsf{I}(s \log p) \log s + \epsilon \,\mathsf{I}(\log p) + s \,\mathsf{I}(\log p) \log \log p\Big).$$

REMARK 8. Let $p_n$ denote the $n$-th prime number. The prime number theorem implies that $p_n \asymp n \log n$. Cramér's conjecture [Cra36] states that

$$\limsup_{n \to \infty} \frac{p_{n+1} - p_n}{(\log p_n)^2} = 1.$$

This conjecture is supported by numerical evidence. Setting

$$N = \max{(2^{l+1}, d)},$$

the conjecture implies that the smallest prime number $p$ with $p > N$ satisfies $p = N + O(\log^2 N)$. Using a polynomial time primality test [AKS04], it follows that this number can be computed by brute force in time $(\log N)^{O(1)}$. In addition, in order to satisfy the complexity bound it suffices to tabulates prime numbers of sizes 2, 4, 8, 16, etc.

### 3.2.2. Chinese remaindering

In our algorithm and Theorem 1, we regard the computation of a prime number $p > N = \max{(2^{l+1}, d)}$ as a precomputation. This is reasonable if $N$ is not too large. Now the quantity $\log d$ usually remains reasonably small. Hence, our assumption that $N$ is not too large only gets violated if $l_P + l_Q$ becomes large. In that case, we will rather use Chinese remaindering. We first compute $r = O(l/\log d)$ prime numbers $p_1 < \cdots < p_r$ with

$$\begin{aligned} p_1 &> d \\ p_1 \cdots p_r &> 2^{l+1}. \end{aligned}$$

Each $\mathbb{F}_{p_k}$ contains a primitive root of unity $\omega_k$ of order $\geqslant d$. We next proceed as before, with $p = p_1 \cdots p_r$ and $\omega \in \mathbb{F}_p = \mathbb{Z}/p\mathbb{Z}$ such that $\omega \bmod p_k = \omega_k$ for each $k$. Indeed, the fact that each $V_{s,X,\omega} \bmod p_k = V_{s,X,\omega_k}$ is invertible implies that $V_{s,X,\omega}$ is invertible.

We will say that $p_1 < \cdots < p_r$ form a reduced sequence of prime moduli with order $d$ and capacity $N$, whenever $p_1 > d$, $p_1 \cdots p_r > N$, $p_1 \cdots p_{r-1} \leqslant N$ and $\log p_r = O(\log d)$. We then have the following refinement of Corollary 7:

COROLLARY 9. *Given $P, Q \in \mathbb{Z}[z_1, ..., z_n]$ and a reduced sequence $p_1 < \cdots < p_r$ of prime moduli with order $d$ and capacity $2^{l+1}$, we can compute $PQ$ in time*

$$O\bigg(\sigma l \log s \, \frac{\mathsf{I}(s \log d)}{s \log d} + \epsilon l \, \frac{\mathsf{I}(\log d)}{\log d} + \sigma \, \mathsf{I}(l) \log l\bigg).$$

## 3.3. Floating point coefficients

An important kind of sparse polynomials are power series in several variables, truncated by total degree. Such series are often used in long term integration of dynamical systems [MB96, MB04], in which case their coefficients are floating point numbers rather than integers. Assume therefore that $P$ and $Q$ are polynomials with floating coefficients with a precision of $\ell$ bits.

Let $e_P$ be the maximal exponent of the coefficients of $P$. For a so called *discrepancy* $\eta_P \in \mathbb{N}$, fixed by the user, we let $\hat{P}$ be the integer polynomial with

$$\hat{P}_i = \lfloor P_i \, 2^{\ell + \eta_P - e_P} \rceil$$

for all $i$. We have $l_{\hat{P}} \leqslant \ell + \eta_P$ and

$$|P - \hat{P}\, 2^{e_P - \ell - \eta_P}| \leqslant 2^{e_P - \ell - \eta_P - 1}$$

for the sup-norm on the coefficients. If all coefficients of $P$ have a similar order of magnitude, in the sense that the minimal exponent of the coefficients is at least $e_P - \eta_P$, then we actually have $P = \hat{P}\, 2^{e_P - \ell - \eta_P}$. Applying a similar decomposition to $Q$, we may compute the product

$$PQ = \hat{P}\hat{Q}\, 2^{e_P + e_Q - 2\ell - \eta_P - \eta_Q}$$

using the algorithm from section 2 and convert the resulting coefficients back into floating point format.

Usually, the coefficients $f_i$ of a univariate power series $f(z)$ are approximately in a geometric progression $\log f_i \sim \alpha\, i$. In that case, the coefficients of the power series $f(\lambda z)$ with $\lambda = e^{-\alpha}$ are approximately of the same order of magnitude. In the multivariate case, the coefficients still have a geometric increase on diagonals $\log f_{\lfloor k_1 i \rfloor, \ldots, \lfloor k_n i \rfloor} \sim \alpha_{k_1, \ldots, k_n}\, i$, but the parameter $\alpha_{k_1, \ldots, k_n}$ depends on the diagonal. After a suitable change of variables $z_i \mapsto \lambda_i z_i$, the coefficients in a big zone near the main diagonal become of approximately the same order of magnitude. However, the discrepancy usually needs to be chosen proportional to the total truncation degree in order to ensure sufficient accuracy elsewhere.

## 3.4. Rational coefficients

Let us now consider the case when $\mathbb{K} = \mathbb{Q}$. Let $q_P$ and $q_Q$ denote the least common multiples of the denominators of the coefficients of $P$ resp. $Q$. One obvious way to compute $PQ$ is to set $\hat{P} := P q_P$, $\hat{Q} := Q q_Q$, and compute $\hat{P}\hat{Q}$ using one of the methods from section 3.2. This approach works well in many cases (e.g. when $P$ and $Q$ are truncations of exponential generating series). Unfortunately, this approach is deemed to be very slow if the size of $q_P$ or $q_Q$ is much larger than the size of any of the coefficients of $PQ$.

An alternative, more heuristic approach is the following. Let $p_1 < p_2 < \cdots$ be an increasing sequence of prime numbers with $p_1 > d$ and such that each $p_i$ is relatively prime to the denominators of each of the coefficients of $P$ and $Q$. For each $i$, we may then multiply $P \bmod p_i$ and $Q \bmod p_i$ using the algorithm from section 2. For $i = 1, 2, 4, 8, \ldots$, we may recover $PQ \bmod p_1 \cdots p_i$ using Chinese remaindering and attempt to reconstruct $PQ$ from $PQ \bmod p_1 \cdots p_i$ using rational number reconstruction [GG02, Chapter 5]. If this yields the same result for a given $i$ and $2i$, then the reconstructed $PQ$ is likely to be correct at those stages.

Of course, if we have an *a priori* bound on the bit sizes of the coefficients of $R$, then we may directly take a sufficient number of primes $p_1 < \cdots < p_r$ such that $R$ can be reconstructed from its reduction modulo $p_1 \cdots p_r$.

## 3.5. Algebraic coefficients

Let $\mathbb{K}$ be an algebraic number field. For some algebraic integer $\alpha$, we may write $\mathbb{K} = \mathbb{Q}[\alpha]$. Let $A \in \mathbb{Z}[u]$ be the monic polynomial of minimal degree $k$ with $A(\alpha) = 0$. Given a prime number $p$, the polynomial $A \bmod p$ induces an algebraic extension $\mathbb{F}_p[\hat{\alpha}]$ of $\mathbb{F}_p$, where $(A \bmod p)(\hat{\alpha}) = 0$. Reduction modulo $p$ of a sparse polynomial $P \in \mathbb{Z}[\alpha][z]$ then yields a sparse polynomial $P \bmod p \in \mathbb{F}_p[\hat{\alpha}][z]$. We have seen in section 3.1 how to multiply sparse polynomials over the finite field $\mathbb{F}_p[\hat{\alpha}]$. Choosing one or more sufficiently large prime numbers $p$, we may thus apply the same approaches as in section 3.2 in order to multiply sparse polynomials over $\mathbb{Z}[\alpha]$. Using the techniques from section 3.4, we next deal with the case of sparse polynomials over $\mathbb{K} = \mathbb{Q}[\alpha]$.

# 4. FAST PRODUCTS OF POWER SERIES

## 4.1. Total degree

Given $i \in \mathbb{N}^n$, let $|i| = i_1 + \cdots + i_n$. The *total degree* of a polynomial $P \in \mathbb{A}[z]$ is defined by

$$\deg P = \max\{|i| : P_i \neq 0\}.$$

Given a subset $I \subseteq \mathbb{N}^n$, we define the *restriction* $P_I$ of $P$ to $I$ by

$$P_I = \sum_{i \in I} P_i z^i.$$

For $d \in \mathbb{N}$, we define initial segments $I_d$ of $\mathbb{N}^n$ by

$$I_d \;=\; \{i \in \mathbb{N}^n : |i| < d\}$$

Then

$$\mathbb{A}[z]_{I_d} = \{P \in \mathbb{A}[z] : \operatorname{supp} P \subseteq I_d\} = \{P_{I_d} : P \in \mathbb{A}[z]\}$$

is the set of polynomials of total degree $< d$. Given $P, Q \in \mathbb{A}[z]_{I_d}$, the aim of this section is to describe efficient algorithms for the computation of $R = (PQ)_{I_d}$. We will follow and extend the strategy described in [LS03].

REMARK 10. *The results of this section generalize* [vdH02a] *to so called pondered total degrees* $|i| = \lambda_1 i_1 + \cdots + \lambda_n i_n$ *with* $\lambda_1, \ldots, \lambda_n > 0$, *but for the sake of simplicity, we will stick to ordinary total degrees.*

## 4.2. Projective coordinates

Given a polynomial $P \in \mathbb{A}[z]$, we define its *projective transform* $\mathrm{T}(P) \in \mathbb{A}[z]$ by

$$\mathrm{T}(P)(z_1, \ldots, z_n) = P(z_1 z_n, \ldots, z_{n-1} z_n, z_n).$$

If $\operatorname{supp} P \subseteq I_d$, then $\operatorname{supp} \mathrm{T}(P) \subseteq J_d$, where

$$J_d \;=\; \{(i_1 + i_d, \ldots, i_{d-1} + i_d, i_d) : i \in I_d\}.$$

Inversely, for any $P \in \mathbb{A}[z]_{J_d}$, there exists a unique $\mathrm{T}^{-1}(P) \in \mathbb{A}[z]_{I_d}$ with $P = \mathrm{T}(\mathrm{T}^{-1}(P))$. The transformation $\mathrm{T}$ is an injective morphism of $\mathbb{A}$-algebras. Consequently, given $P, Q \in \mathbb{A}[z]_{I_d}$, we will compute the truncated product $(PQ)_{I_d}$ using

$$(PQ)_{I_d} = \mathrm{T}^{-1}((\mathrm{T}(P)\, \mathrm{T}(Q))_{J_d}).$$

Given a polynomial $P \in \mathbb{A}[z]$ and $j \in \mathbb{N}$, let

$$P_j = \sum_{i_1, \ldots, i_{n-1}} P_{i_1, \ldots, i_{n-1}, j}\, z_1^{i_1} \cdots z_{n-1}^{i_{n-1}} z_n^j \in \mathbb{Z}[z_1, \ldots, z_{n-1}].$$

If $\operatorname{supp} P \subseteq J_d$, then $\operatorname{supp} P_j \subseteq X$, with

$$X \;=\; \{i \in \mathbb{N}^{n-1} : i_1 + \cdots + i_{n-1} < d\}.$$

## 4.3. Multiplication by evaluation and interpolation

Let $\omega$ be an element of $\mathbb{K}$ of sufficiently high order $\geqslant d^{n-1}$. Taking $X$ as above, the construction in section 2.3 yields a $\mathbb{K}$-linear and invertible evaluation mapping

$$\mathrm{E} : \mathbb{A}[z]_X \longrightarrow \mathbb{A}^X,$$

such that for all $P, Q \in \mathbb{A}[z]_X$ with $PQ \in \mathbb{A}[z]_X$, we have

$$\mathrm{E}(PQ) = \mathrm{E}(P)\, \mathrm{E}(Q). \tag{1}$$

This map extends to $\mathbb{A}[z]_X[z_n]$ using

$$\mathrm{E}(P_0 + \cdots + P_k z^n) \;=\; \mathrm{E}(P_0) + \cdots + \mathrm{E}(P_k)\, z_n^k \in \mathbb{A}^X[z_n].$$

Given $P, Q \in \mathbb{A}[z]_{J_d}$ and $j < d$, the relation (1) yields

$$\mathrm{E}((PQ)_j) = \mathrm{E}(P_j)\, \mathrm{E}(Q_0) + \cdots + \mathrm{E}(P_0)\, \mathrm{E}(Q_j).$$

In particular, if $R = (PQ)_{J_d}$, then

$$\mathrm{E}(R) = \mathrm{E}(P)\, \mathrm{E}(Q) \bmod z_n^d.$$

Since E is invertible, this yields an efficient way to compute $R$.

## 4.4. Complexity analysis

The number of coefficients of a truncated series in $\mathbb{A}[z]_{I_d}$ is given by

$$|I_d| \;=\; \binom{n+d-1}{n}$$

The size $s = |X|$ of $X$ is smaller by a factor between 1 and $d$:

$$s \;=\; \binom{n+d-2}{n-1} = \frac{n}{n+d-1} |I_d|.$$

THEOREM 11. *Given $P, Q \in \mathbb{A}[z]_{I_d}$ and an element $\omega \in \mathbb{K}$ of order at least $d^{n-1}$, we can compute $(PQ)_{I_d}$ using $O(s\, d)$ inversions in $\mathbb{K}$, $O(s\, \mathsf{M}(d))$ general operations in $\mathbb{A}$, and $O(d\, \mathsf{M}(s)\log s)$ vectorial operations in $\mathbb{A}$.*

PROOF. The transforms T and $\mathrm{T}^{-1}$ require a negligible amount of time. The computation of the evaluation points $\omega^{k_i}$ only involves $O(s)$ products in $\mathbb{K}$, when exploiting the fact that $X$ is an initial segment. The computation of $\mathrm{E}(\mathrm{T}(P))$ and $\mathrm{E}(\mathrm{T}(Q))$ requires $O(d\, \mathsf{M}(s)\log s)$ vectorial operations in $\mathbb{A}$. The computation of $\mathrm{E}(\mathrm{T}(P))\, \mathrm{E}(\mathrm{T}(Q)) \bmod z_n^d$ can be done using $O(s\, \mathsf{M}(d))$ general operations in $\mathbb{A}$. Recovering $R$ again requires $O(d\, \mathsf{M}(s)\log s)$ vectorial operations in $\mathbb{A}$, as well as $O(s\, d)$ divisions in $\mathbb{K}$. □

COROLLARY 12. *Given $P, Q \in \mathbb{F}_p[z]_{I_d}$, where $p$ is a prime number with $p-1 \geqslant d^{n-1}$, we can compute $(PQ)_{I_d}$ in time $O(\mathsf{I}(s\, d\log p)\log s + s\, d\, \mathsf{I}(\log p)\log\log p)$.*

In the case when $P, Q \in \mathbb{Z}[z]_{I_d}$, the assumption $p > \max(2^{l+1}, d^{n-1})$, with $l = l_P + l_Q + \log_2 s$, guarantees that the coefficients of the result $PQ$ can be reconstructed from their reductions modulo $p$. Combining this observation with Chinese remaindering, we obtain:

COROLLARY 13. *Given $P, Q \in \mathbb{Z}[z]_{I_d}$ and a reduced sequence $p_1 < \cdots < p_r$ of prime moduli with order $d^{n-1}$ and capacity $2^{l+1}$, we can compute $(PQ)_{I_d}$ in time*

$$O\!\left(\frac{\mathsf{I}(s\, d\, n\log d)}{n\log d}\, l\log s + \frac{\mathsf{I}(n\log d)}{n\log d}\, l\, s\, d\log n\log\log d + |I_d|\, \mathsf{I}(l)\log l\right).$$

## 5. Implementation and timings

We have implemented the fast series product of the previous section within the C++ library multimix of Mathemagix [vdH+02b]. We report on timings for when $\mathbb{K} = \mathbb{F}_p$, with $p = 3.2^{30} + 1$, on a 2.4 GHz Intel(R) Core(TM)2 Duo platform. Recall that $n$ is the number of the variables and $d$ the truncation order. Timings are given in milliseconds. The line *naive* corresponds to the naive multiplication, that performs all the two by two monomial products, while the line *fast* stands for the algorithm of the previous section. We have added the size $|I_d|$ of the support of the series, together with the cost $\mathsf{M}(|I_d|)$ of our univariate multiplication in size $|I_d|$ for comparison. An empty cell corresponds to a computation that needed more than 10 minutes. The following tables demonstrate that the new fast algorithms are relevant to practice and that the theoretical *softly linear* asymptotic cost can really be observed.

| $d$ | 12 | 22 | 42 | 82 | 162 |
|---|---|---|---|---|---|
| naive | 0.4 | 3.6 | 41 | 578 | 10119 |
| fast | 3 | 12 | 66 | 322 | 1509 |
| $|I_d|$ | 78 | 253 | 903 | 3403 | 13203 |
| $\mathsf{M}(|I_d|)$ | 0.1 | 0.4 | 1.6 | 7.4 | 28 |

**Table 1.** Series product with 2 variables

| $d$ | 12 | 22 | 42 |
|---|---|---|---|
| naive | 50 | 4630 | |
| fast | 172 | 2345 | 45702 |
| $|I_d|$ | 1365 | 12650 | 148995 |
| $\mathsf{M}(|I_d|)$ | 3 | 28 | 902 |

**Table 2.** Series product with 4 variables

| $d$ | 7 | 12 | 17 | 22 | 27 |
|---|---|---|---|---|---|
| naive | 25 | 5346 | | | |
| fast | 159 | 3775 | 30773 | 168873 | 612310 |
| $|I_d|$ | 924 | 12376 | 74613 | 296010 | 906192 |
| $\mathsf{M}(|I_d|)$ | 1.6 | 28 | 243 | 886 | 2887 |

**Table 3.** Series product with 6 variables